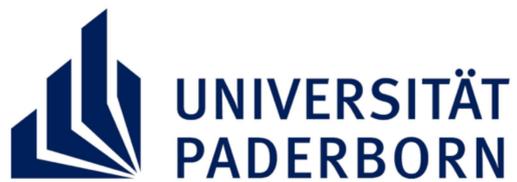

Thema

# Eine empirische Analyse der Skalierung von Value-at-Risk Schätzungen

Bachelorarbeit zur Erlangung des akademischen Grades
Bachelor of Science – Wirtschaftswissenschaften

vorgelegt von: Kuhlmann, Marita

Matrikelnummer: 7081983

Paderborn, den 21.01.2020

# Inhaltsverzeichnis

# Abbildungsverzeichnis

# Tabellenverzeichnis

# Abkürzungs- und Symbolverzeichnis

## Abkürzungen

| | |
|---|---|
| BCOM | Bloomberg Commodity Index |
| DAX | Deutscher Aktienindex |
| EURUSD | Euro/US-Dollar Kurs |
| GARCH | Generalized Autoregressive Conditional Heteroscedasticity |
| GSCI | Goldman Sachs Commodity Index |
| JPYEUR | japanischer Yen/Euro Kurs |
| S&P | Standard & Poor's |
| VaR | Value-at-Risk |

## Symbole

| | |
|---|---|
| $K_t$ | Kurs/Preis in t |
| $LR_{cc}$ bzw. $LR_{uc}$ | Log-Likelihood-Ratio Teststatistik des conditional bzw. unconditional coverage Backtests |
| p | Konfidenzniveau |
| $r_{t,1}$ bzw. $r_{t,10}$ | Logarithmierte eintägige bzw. zehntägige Rendite im Zeitpunkt t |
| $t_{1-p,v}$ | Quantil der t-Verteilung (mit $v$ Freiheitsgraden) |
| T | Haltedauer des VaR |
| $VaR_{1-p}(T)$ | T-Tages VaR |
| $z_{1-p}$ | Quantil der Standardnormalverteilung |
| $\mu_t$ | Erwartungswert/Mittelwert der Renditen im Zeitpunkt t |
| $\sigma_t$ | Standardabweichung/Volatilität der Renditen im Zeitpunkt t |
| $v$ | Anzahl der Freiheitsgrade |

# 1. Einleitung

Der Value-at-Risk (VaR) hat sich wegen verschiedener Vorteile als ein Standardrisikomaß etabliert und gilt als relevant für die Industrie. Auch wenn anstelle des VaR heutzutage vermehrt auf den Expected Shortfall als eine Alternative zurückgegriffen wird, findet der VaR beispielsweise immer noch Anwendung beim Backtesting in der Vorschrift „Basel III" sowie zur Bestimmung von Ausfallrisiken (vgl. BCBS 2011, S. 17, S. 2; BCBS 2019, S. 81, S. 94; BCBS 2017, S. 67).

In der Praxis wird zur Bestimmung des VaR für eine längere Haltedauer häufig auf die Skalierung mit der „Square Root of Time Rule" oder Quadratwurzel-T-Regel zurückgegriffen. Dabei wird der VaR für eine kürzere Haltedauer bestimmt und dann der eigentlich gewünschten Haltedauer entsprechend hochskaliert. So ist es beispielsweise in den Basler Vorschriften erlaubt, dass Banken zur Bestimmung des 10-Tages VaR den 1-Tages VaR mit der Wurzel aus zehn hochskalieren (vgl. Alexander 2008, S. 21; Basler Ausschuss für Bankenaufsicht 2005, S. 46).

Angewendet werden kann diese Regel jedoch nur unter bestimmten Voraussetzungen, sodass dann idealerweise die direkte und skalierte Schätzung des VaR für die zugrundeliegende Haltedauer identische Ergebnisse liefern (vgl. Meyer 1999, S. 360).

Tatsächlich ließ sich aber in den verschiedensten Untersuchungen dieser Skalierungsregel feststellen, dass genau dies nicht zu beobachten ist:

J.P. Morgan und Reuters (1996, S. 87 f.) weisen darauf hin, dass durch die Skalierung unter anderem bei Währungskursen und Aktienindizes die Schätzergebnisse sehr unwahrscheinlich und unsinnig sind. Grund dafür ist, dass der Faktor zur Skalierung die realen Bedingungen nicht mit einbezieht.

Danielsson und Zigrand (2003) kommen zu dem Entschluss, dass durch die Skalierung mit der Quadratwurzel-T-Regel das Risiko unterschätzt wird. Diese Verzerrung wird insbesondere für längere Halteperioden und niedrigere Quantile umso größer.

Auch die Ergebnisse von Cameron et al. (2016) deuten darauf hin, dass die Skalierung der Volatilität mit $\sqrt{10}$ auf zehn Tage in einer erheblichen Unterschätzung des VaR resultiert, da der Skalierungsfaktor zu gering ist.

Mit unterschiedlichen Eigenschaften der Renditen, die mögliche Gründe für die Verzerrung sein könnten, beschäftigen sich Wang et al. (2011). Insbesondere bei einer seriellen Abhängigkeit der Daten und Verteilungen mit breiten Enden („fat tails") ist eine bedeutende Verzerrung festzustellen, wobei breite Enden zu einer Überschätzung des VaR führen. Im Durchschnitt werden der 10-Tages und 30-Tages VaR durch die Skalierung unterschätzt, wodurch die Anwendung dieser Regel zu ungeeigneten Kapitalanforderungen führt.

Bei Untersuchungen zur Skalierung der bedingten Volatilität stellen Saadi und Rahman (2008) fest, dass bei der Anwendung von GARCH-Modellen die Volatilität unterschätzt wird.

Für ein GARCH(1,1)-Modell finden hingegen Diebold et al. (1998) heraus, dass die Ergebnisse bei der Skalierung im Durchschnitt einwandfrei sind, auch wenn die Voraussetzungen nicht erfüllt werden. Deshalb sei es verständlich, dass diese Regel häufig Anwendung findet, aber es sollte dennoch eine bessere Alternative gefunden werden.

Der Basler Ausschuss selbst merkt in einem Arbeitspapier an, dass nach unterschiedlichen Forschungen die Skalierung mit der Quadratwurzel-T-Regel falsche Schätzergebnisse liefert. Weil sich jedoch keine sofortige Alternative für diese Regel finden lässt, findet sie immer noch Anwendung, besonders da sie sehr nützlich ist (vgl. BCBS 2011, S. 1, S. 8). Zur Validierung der verwendeten VaR-Schätzmodelle wird in den Vorschriften jedoch auch gefordert, dass zusätzlich die getroffenen Annahmen auf ihre Adäquanz überprüft werden sollen. Darunter zählt beispielsweise auch die Anwendung der Quadratwurzel-T-Regel (vgl. Basler Ausschuss für Bankenaufsicht 2005, S. 55).

Ziel dieser Arbeit ist es nun, für verschiedene VaR-Schätzmethoden und Datenkategorien die Schätzgenauigkeit und Anwendbarkeit der Quadratwurzel-T-Regel zu prüfen. Dafür soll die direkte Schätzung des VaR mit der skalierten Schätzung verglichen werden. Untersucht wird der 10-Tages VaR für die Konfidenzniveaus 95% und 99%. Anhand von zwei Backtest-Methoden werden die Schätzergebnisse ausgewertet. Die gesamte Analyse wird mithilfe der Statistiksoftware *R* durchgeführt.

In dem folgenden Teil dieser Arbeit werden zunächst der VaR und die Skalierung vorgestellt. Daraufhin wird in Kapitel 3 auf die verwendete Methodik zur VaR-Schätzung und zum Backtest eingegangen. Als VaR-Schätzmethoden werden die historische Simulation und die Varianz-Kovarianz-Methode vorgestellt und kurz hinsichtlich ihrer Vor- und Nachteile beurteilt. Danach wird auf den unconditional coverage und conditional coverage Backtest eingegangen. Im vierten Kapitel erfolgt die empirische Analyse, in der zum einen die Daten vorgestellt werden, das Vorgehen bei der Schätzung des VaR erläutert wird und die Backtest-Ergebnisse ausgewertet werden. Eine Diskussion der Ergebnisse folgt in Kapitel 5. Zum Schluss werden die Ergebnisse in einem Fazit zusammengefasst.

## 2. Value-at-Risk

### 2.1. Definition

Der Value-at-Risk ist definiert als der Verlust, der mit einer gegebenen Wahrscheinlichkeit (auch: Konfidenzniveau) in einer bestimmten Haltedauer nicht überschritten wird (vgl. Priermeier 2005, S. 51).

Zur Schätzung des VaR sollten somit vorab das Konfidenzniveau und die Haltedauer festgelegt werden. Die Höhe des VaR wird dabei maßgebend durch die Wahl dieser beiden Parameter bestimmt. Grundsätzlich gilt, dass der VaR für ein höheres Konfidenzniveau c.p. größer ist und ebenso für eine längere Haltedauer c.p. ein größerer VaR geschätzt wird. Üblicherweise werden Konfidenzniveaus von mindestens 95% gewählt. Durch die Bankenaufsicht wird ein 99%iges Konfidenzniveau vorgegeben, aber bei der Verwendung von internen Modellen liegt das Konfidenzniveau meist zwischen 95% und 99% (vgl. Meyer 1999, S. 94, 97 f.).

Der wohl größte Kritikpunkt an dem VaR ist, dass es sich wegen der fehlenden Eigenschaft der Subadditivität um kein kohärentes Risikomaß handelt. Hinzu kommt, dass die Verluste, die den VaR überschreiten, unberücksichtigt bleiben. Der Expected Shortfall hingegen kompensiert genau diese Nachteile des VaR, weshalb dieser auch nun vermehrt den VaR ersetzt. Gegenüber dem Expected Shortfall bleibt der VaR jedoch in der Hinsicht vorteilhafter, dass er praktisch einfacher anzuwenden ist und sich auch das Backtesting einfacher gestaltet (vgl. BCBS 2011, S. 18 ff.; McNeil, Frey und Embrechts 2015, S. 76 f.).

### 2.2. Skalierung

Für die Anwendung der Quadratwurzel-T-Regel wird vorausgesetzt, dass es sich um eine lineare Risikoposition handelt und die Renditen unabhängig identisch sowie normal verteilt sind (vgl. J.P.Morgan/Reuters 1996, S. 37; Alexander 2008, S. 21).

Während der Erwartungswert und die Varianz unter diesen Voraussetzungen linear mit der Zeit wachsen, wächst die Standardabweichung dementsprechend mit der Quadratwurzel der Zeit (vgl. Jorion 2001, S. 103). Somit ergibt sich als Formel für den VaR mit dem Konfidenzniveau p und einer Haltedauer T:

$$VaR_{1-p}(T) = VaR_{1-p}(1) \cdot \sqrt{T}, \tag{1}$$

wobei $VaR_{1-p}(1)$ den 1-Tages VaR kennzeichnet. Zudem wird vereinfachend angenommen, dass der Mittelwert der Renditen 0 beträgt (vgl. Johanning 1998, S. 31; Tsay 2013, S. 339).

In Anbetracht der Charakteristiken, die die Renditeverteilung in der Realität aufweist, gelten die Annahmen der Quadratwurzel-T-Regel als unrealistisch. Im Allgemeinen ist zu beobachten, dass Renditen nicht unabhängig identisch verteilt sind und auch nicht einer Normalverteilung folgen. Vielmehr weisen Renditen Verteilungen mit breiteren Enden („fat tails") auf, d.h. extreme Werte treten häufiger ein als bei der Normalverteilung (vgl. McNeil, Frey und Embrechts 2015, S. 79 f., S. 85).

# 3. Methodik

In diesem Kapitel soll in einem ersten Teil auf die für diese Arbeit relevanten Schätzungsmethoden des VaR eingegangen werden. Einer kurzen Beschreibung folgend werden die Verfahren hinsichtlich ihrer Vor- und Nachteile beurteilt. Der zweite Teil dieses Kapitels stellt das Backtesting sowie ebenfalls dessen verwendete Methoden vor.

## 3.1. Value-at-Risk Schätzmethoden

Grundsätzlich sind zur Bestimmung des VaR zwei Wege zu unterscheiden: parametrische Ansätze sowie nichtparametrische (oder auch simulative) Ansätze. Im Grunde des Vorgehens sind beiden Ansätzen die Schätzung der Wahrscheinlichkeitsverteilung der Wertänderungen gemeinsam, um letztendlich den VaR als Quantil dieser Verteilung zu bestimmen (vgl. Johanning 1998, S. 24). Während für die parametrischen Verfahren eine Annahme für die Verteilung der Renditen getroffen bzw. unterstellt werden muss, gehen die simulativen Ansätze direkt von der aus den historischen Daten beobachteten Verteilung aus (vgl. Priermeier 2005, S. 51; Meyer 1999, S. 125 f.).

In dieser Arbeit wird als simulativer Ansatz die historische Simulation verwendet und als parametrischer Ansatz die Varianz-Kovarianz-Methode. Beide zählen zu den häufig verwendeten Methoden bei der VaR-Schätzung und sollen aus diesem Grund wegen ihres Praxisbezuges näher untersucht werden (vgl. Committee of European Securities Regulators 2010, S. 28).

### 3.1.1. Historische Simulation

Bei der historischen Simulation wird zur Schätzung des VaR die Verteilung der Gewinne und Verluste der Position verwendet, welche mithilfe der historisch beobachteten Preisveränderungen bestimmt wird. Die Gewinne und Verluste der Größe nach aufsteigend geordnet, kann der VaR anschließend schlicht durch Abzählen bestimmt werden: eine durch das Quantil des Konfidenzniveaus ermittelte Anzahl der größten Verluste wird dabei gedanklich gestrichen. Der darauf nach Erreichen des Quantils höchste verbleibende Verlust stellt den VaR dar (vgl. Meyer 1999, S. 192; Priermeier 2005, S. 54 f.).

Voraussetzungen zur Anwendung der historischen Simulation sind stationäre sowie unabhängig identisch verteilte Renditen in dem betrachteten Zeitraum und die Linearität der Risikoposition (vgl. Johanning 1998, S. 32; Meyer 1999, S. 196).

Die Vorteile dieser Methode liegen zum einen in der einfachen Implementierbarkeit und leichten Bestimmung des VaR. Des Weiteren wird keine spezielle Verteilungsannahme der Renditen unterstellt, weshalb die Methode auch auf eine Vielzahl von Verteilungen angewendet werden kann. Somit können auch breite Enden einer Renditeverteilung berücksichtigt werden. Da das Vorgehen und die Resultate ingesamt intuitiv und schlüssig sind, ist das Verfahren zudem leicht zu erklären (vgl. Jorion 2001, S. 222 f.; Meyer 1999, S. 197).

Ein Nachteil der historischen Simulation ist, dass eine ausreichend große Historie an Daten gebraucht wird bzw. vorhanden sein sollte (vgl. Jorion 2001, S. 223). Vor dem Hintergrund, dass der geschätzte VaR nur innerhalb der bereits beobachteten historischen Verteilung liegen kann, ist auch die Auswahl der Länge des Zeitraums nicht trivial. Es wird angenommen, dass die Ereignisse und Verläufe der Wertänderungen in der Vergangenheit ebenso relevant für die Zukunft sind. Dies ist aber bei einem längeren Zeitraum nicht immer der Fall, da sich unter Umständen die Marktbedingungen im Zeitverlauf bedeutend geändert haben könnten. Bei einem kurzen Zeitraum kann es, je nachdem, ob dieser mehr bzw. weniger volatil ist, zu einer Über- bzw. Unterschätzung des Risikos kommen. Zudem sind bei einem kleinen Stichprobenumfang Schätzfehler wahrscheinlicher. So kann es in diesen Fällen durch die Wahl des Zeitraums zu Verzerrungen kommen (vgl. Johanning 1998, S. 34; Jorion 2001, S. 223; Meyer 1999, S. 198).

### 3.1.2. Varianz-Kovarianz Methode

Um den VaR mit der Varianz-Kovarianz Methode zu bestimmen, ist es im Grunde nur notwendig, den Erwartungswert ($\mu_t$) und die Standardabweichung bzw. Volatilität ($\sigma_t$) der Renditen zu schätzen (vgl. Meyer 1999, S. 134). Häufig wird für die Renditen eine Normalverteilung angenommen. Mit dem Quantil der Standardnormalverteilung ($z_{1-p}$) und den geschätzten Parametern kann der VaR folgendermaßen bestimmt werden:

$$VaR_{1-p} = \mu_t + z_{1-p}\sigma_t. \qquad (2)$$

Analog kann unter Annahme der t-Verteilung der VaR mit der Formel

$$VaR_{1-p} = \mu_t + t_{1-p,v}\sigma_t \qquad (3)$$

bestimmt werden, wobei $t_{1-p,v}$ das Quantil der t-Verteilung mit $v$ Freiheitsgraden darstellt (vgl. Tsay 2013, S. 332). Die Verwendung der t-Verteilung ist insbesondere für Verteilungen mit breiten Enden sehr üblich (vgl. Daníelsson und Zigrand 2003, S. 3).

Als Erweiterung der Methode kann auch die bedingte Standardabweichung mit GARCH-Modellen geschätzt und bei der Ermittlung des VaR verwendet werden (vgl. Meyer 1999, S. 138). Dadurch kann die Volatilität zeitabhängig modelliert und die breiten Enden einer Verteilung berücksichtigt werden (vgl. Meyer 1999, S. 114).

Die Varianz-Kovarianz Methode kann einfach und auch schnell umgesetzt werden. Probleme könnten sich bei der Annahme der Normalverteilung ergeben, da so wegen der breiten Enden der Renditeverteilung das Risiko unterschätzt wird. Des Weiteren kann die Methode nur für wenige parametrische Verteilungen, wie die Normal- oder t-Verteilung, verallgemeinert werden und nur für lineare Positionen adäquat angewendet werden (vgl. Jorion 2001, S. 220 f.; Alexander 2008, S. 51).

### 3.2. Backtesting

Das Backtesting stellt zur Überprüfung des Modells hinsichtlich seiner Annahmen und Parameter eine wichtige Rolle dar, wenn es zur Anwendung des VaR in der Praxis kommt. Im Rahmen der Basler Vorschriften für das Eigenkapital von Banken ist das Backtesting insbesondere bei der Nutzung von internen VaR-Modellen wichtig, um auf eine Über- oder Unterschätzung des Risikos zu prüfen (vgl. Jorion 2001, S. 129).

Grundsätzlich kann je nach Konfidenzniveau eine erwartete Anzahl an Überschreitungen bestimmt werden. So wird nach der Definition des VaR bei einem Konfidenzniveau von beispielsweise 95% erwartet, dass nicht mehr als 5% der Beobachtungen den VaR überschreiten. Bei dem Backtesting wird anhand der auftretenden Gewinne und Verluste geprüft, wie häufig die VaR-Schätzungen tatsächlich überschritten werden (vgl. Jorion 2011, S. 358).

Dabei gelten, verglichen mit der erwarteten Anzahl, zu viele Überschreitungen als problematisch, da das Risiko unterschätzt wird und zu wenig Kapital zur Deckung der Risiken vorhanden ist. Aber auch zu wenige Überschreitungen des VaR sind ein Problem, da solche VaR-Schätzungen zu einer ineffizienten Verteilung des Kapitals führen (vgl. Jorion 2001, S. 130).

Als Backtest Methoden werden in dieser Arbeit der unconditional coverage Test von Kupiec und der conditional coverage Test von Christoffersen angewendet. Bei beiden wird davon ausgegangen, dass die Überschreitungen als Zufallsvariable einem Bernoulliprozess folgt, also jeweils die Zahl „0" bei keiner Überschreitung oder „1" bei Überschreitung des VaR annimmt (vgl. Jorion 2011, S. 358). Anhand der Anzahl der Überschreitungen wird eine Log-Likelihood Ratio (kurz: LR) als Teststatistik ermittelt, mit der die zu untersuchende Nullhypothese ausgewertet wird. Die Nullhypothese wird abgelehnt, wenn die Teststatistik einen gewissen kritischen Wert überschreitet.

Bei dem unconditional coverage Test wird als Nullhypothese geprüft, ob die Anzahl der Überschreitungen des VaR der erwarteten Anzahl an Überschreitungen entspricht bzw. nicht zu sehr davon abweicht. Der kritische Wert bestimmt sich dabei durch die Chi-Quadrat-Verteilung mit einem Freiheitsgrad, da die LR-Teststatistik asymptotisch dieser Verteilung folgt.

Als Erweiterung des unconditional coverage Test wird bei dem conditional coverage Test zusätzlich noch geprüft, ob die Überschreitungen voneinander unabhängig sind, also nicht in einer Zeit besonders gehäuft auftreten. Hier bestimmt sich der kritische Wert zur Auswertung der Nullhypothese ebenfalls durch die Chi-Quadrat-Verteilung, jedoch mit zwei Freiheitsgraden (vgl. Alexander 2008, S. 337-340).

Um auch hier den Bezug zu der Praxis bzw. insbesondere dem Baseler Rahmenwerk zu behalten, wurden genau diese beiden Backtest-Methoden ausgewählt. Ebenso wie der unconditional und conditional coverage Test, beruht auch das Drei-Zonen-Konzept als Backtest-Methode des Basler Ausschusses auf der Anzahl der Überschreitungen. Da sich dieses auf einen Backtest-Zeitraum von 250 Beobachtungen bezieht, aber in dieser Arbeit ein längerer Zeitraum zum Backtesting betrachtet wird, wird entsprechend auf die coverage Tests zurückgegriffen (vgl. Basler Ausschuss für Bankenaufsicht 1996, S. 5 f.; BCBS 2019, S. 82).

In Zeiten sich ständig verändernder Marktbedingungen wird der conditional coverage Backtest besonders relevant. Denn wenn die Überschreitungen nicht unabhängig sind, wird deutlich, dass das Modell nicht flexibel genug auf solche Veränderungen reagiert (vgl. Jorion 2001, S. 142; Alexander 2008, S. 338). Auch der Basler Ausschuss weist auf die Wichtigkeit von unabhängigen Überschreitungen hin (vgl. Basler Ausschuss für Bankenaufsicht 1996, S. 5).

# 4. Empirische Analyse

In diesem Kapitel wird die zuvor vorgestellte Methodik nun angewendet. Zunächst wird auf die Datenauswahl sowie eine deskriptive Statistik dieser eingegangen. Daraufhin folgt die Schätzung des VaR für eine Haltedauer von 10 Tagen und den Konfidenzniveaus 95% und 99% nach der direkten und skalierten Ermittlung. Mit dem Backtesting werden die Ergebnisse anschließend ausgewertet.

## 4.1. Daten

Untersucht wurden je zwei Datensätze aus drei unterschiedlichen Kategorien, darunter Aktienindizes, Rohstoffindizes und Währungskurse.

Als Aktienindizes wurden der DAX 30 und S&P 500 gewählt. Die untersuchten Rohstoffindizes sind der S&P GSCI sowie der Bloomberg Commodity Index (kurz: BCOM) und die beiden Währungskurse sind der Euro/US-Dollar Kurs (EURUSD) und der japanische Yen/Euro Kurs (JPYEUR). Die Daten umfassen einen Zeitraum von 10 Jahren (01.11.2009 – 01.11.2019) und stammen aus der Datenbank „Thomson Reuters Eikon".

Von den Preisen und Kursen ($K_t$) ausgehend wurden die logarithmierten Renditen für einen Tag und für zehn Tage berechnet. Die täglichen Renditen berechnen sich wie folgt (vgl. Meyer 1999, S. 22):

$$r_{t,1} = \ln K_t - \ln K_{t-1} = \ln \frac{K_t}{K_{t-1}}. \tag{4}$$

Analog gilt für die 10-Tages Renditen dementsprechend:

$$r_{t,10} = \ln K_t - \ln K_{t-10} = \ln \frac{K_t}{K_{t-10}}. \tag{5}$$

In *R* erfolgt die Berechnung der Renditen mit dem *diff(log)*-Befehl. Für die 10-Tages Renditen wird zusätzlich ein Abstand der Renditen von 10 angegeben, damit anstelle des Vortageskurs ($K_{t-1}$) der Kurs vor 10 Tagen ($K_{t-10}$) verwendet wird.

In den Tabellen 1 und 2 sind für die 1-Tages und 10-Tages Renditen die Ergebnisse der deskriptiven Statistik zusammengefasst.

Tabelle 1: Deskriptive Statistik der 1-Tages Renditen

|  | DAX | S&P | GSCI | BCOM | EURUSD | JPYEUR |
|---|---|---|---|---|---|---|
| Anzahl der Beobachtungen | 2610 | 2610 | 2610 | 2610 | 2610 | 2610 |
| Minimum [in %] | -7,067 | -6,896 | -6,743 | -4,783 | -2,601 | -5,498 |
| Maximum [in %] | 5,210 | 4,840 | 7,617 | 3,702 | 2,259 | 3,925 |
| Mittelwert [in %] | 0,033 | 0,042 | -0,022 | -0,017 | 0,011 | -0,004 |
| Standardabweichung [in %] | 1,177 | 0,920 | 1,190 | 0,843 | 0,552 | 0,677 |
| Schiefe | -0,2859 | -0,4910 | -0,1722 | -0,2376 | -0,0273 | -0,0677 |
| Kurtosis | 5,7454 | 7,7414 | 5,7563 | 5,2499 | 4,5864 | 7,3018 |

Tabelle 2: Deskriptive Statistik der 10-Tages Renditen

|  | DAX | S&P | GSCI | BCOM | EURUSD | JPYEUR |
|---|---|---|---|---|---|---|
| Anzahl der Beobachtungen | 2601 | 2601 | 2601 | 2601 | 2601 | 2601 |
| Minimum [in %] | -25,621 | -17,790 | -14,875 | -11,621 | -5,726 | -8,953 |
| Maximum [in %] | 11,925 | 9,074 | 10,384 | 9,160 | 7,133 | 7,914 |
| Mittelwert [in %] | 0,326 | 0,401 | -0,233 | -0,178 | 0,112 | -0,040 |
| Standardabweichung [in %] | 3,562 | 2,609 | 3,775 | 2,608 | 1,713 | 2,121 |
| Schiefe | -0,9244 | -1,0023 | -0,4192 | -0,3425 | 0,1357 | 0,0073 |
| Kurtosis | 6,5067 | 6,3121 | 3,2781 | 3,7488 | 3,6207 | 3,9172 |

Mit dem Jarque-Bera-Test wurde anhand der Schiefe und Kurtosis getestet, ob eine Normalverteilung für die Datensätze vorliegt. Die Schiefe, welche die Symmetrie der Verteilung angibt, sollte bei Vorliegen einer Normalverteilung gleich 0 sein. Anhand der Kurtosis wird gemessen, ob die Verteilung breite Enden aufweist, wobei für die Normalverteilung ein Wert von 3 angenommen wird. Bei Werten von größer als 3 liegen breite Enden vor, d.h. die Verteilung ist leptokurtisch (vgl. Tsay 2013, S. 22-26). In *R* wird der Jarque-Bera-Test mit dem Befehl *normalTest* aus dem Paket *fBasics* durchgeführt.

Für die 1-Tages und 10-Tages Renditen wurde bei allen Datensätzen nach der Jarque-Bera-Teststatistik die Normalverteilungshypothese abgelehnt. Insbesondere für die 1-Tages Renditen lässt sich anhand der Kurtosis beobachten, dass eine leptokurtische Verteilung vorliegt. Bei den 10-Tages Renditen sind in dieser Hinsicht die Aktienindizes besonders auffällig, die im Vergleich zu den anderen Datensätzen höhere Kurtosis-Werte aufweisen.

### 4.2. Value-at-Risk Schätzungen

Die Schätzung des VaR erfolgt rollierend für eine Fenstergröße von 250 Beobachtungen, was ungefähr einem Handelsjahr entspricht. So werden für die rollierende Schätzung genügend und auch immer noch relevante Daten mit einbezogen. Zur Schätzung des ersten VaR Wertes werden die ersten 250 Beobachtungen verwendet. Für die weiteren Schätzungen wird das Fenster im Zeitverlauf jeweils um einen Tag weitergeschoben. Mit der rollierenden Schätzung wird berücksichtigt, dass sich mit der Zeit die Märkte entwickeln und die Risiken verändern (vgl. Meyer 1999, S. 195). In *R* wird die rollierende Schätzung mit der Funktion *rollapply* aus dem Paket *zoo* durchgeführt.

Für die Schätzung des 10-Tages VaR nach der direkten Ermittlung werden die zehntägigen Renditen verwendet und für die skalierte Ermittlung die eintägigen Renditen zur Bestimmung des 1-Tages VaR mit anschließender Skalierung.

Die Annahme zur Skalierung gemäß der Formel 6, dass der Mittelwert der 1-Tages Renditen 0 beträgt, wurde anhand eines t-Tests für alle Datensätze überprüft und bestätigt. Dementsprechend wird für die Schätzung des VaR nach der Skalierung der Mittelwert nicht miteinbezogen. Da hingegen der Mittelwert der 10-Tages Renditen signifikant von 0 abweicht, wird dieser bei der direkten Ermittlung in der Schätzung berücksichtigt.

Für die historische Simulation wird aus dem Paket *PerformanceAnalytics* der Befehl *VaR* verwendet. Die Abbildungen 1 und 2 zeigen grafisch die hiernach geschätzten VaR-Werte für ein Konfidenzniveau von 95% und 99% für den DAX. In grau sind die zehntägigen Renditen abgebildet. Die rote Linie repräsentiert die Schätzung nach der direkten Methode und die blaue Linie die Schätzung nach der Skalierung.

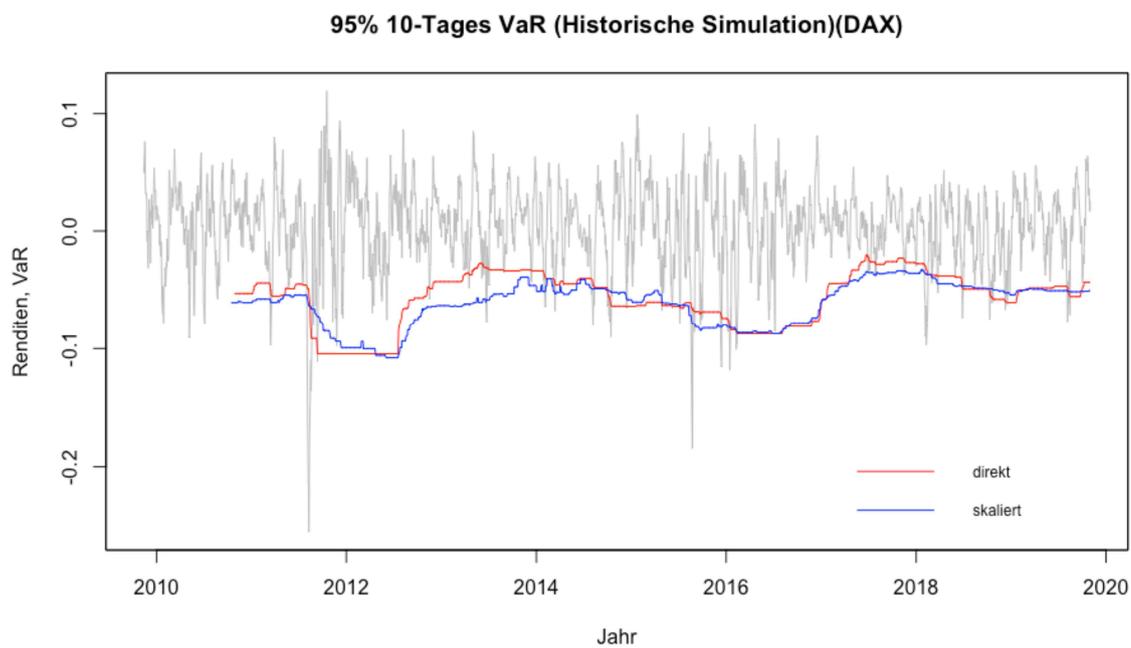

Abbildung 1: 95% 10-Tages VaR (Historische Simulation, DAX)

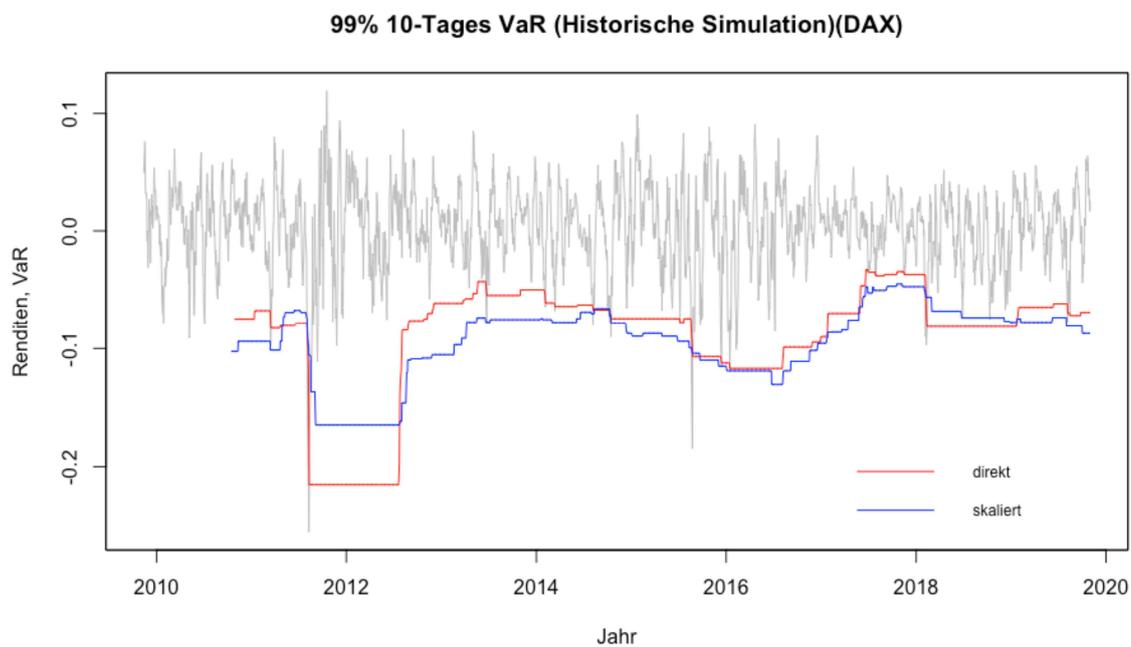

Abbildung 2: 99% 10-Tages VaR (Historische Simulation, DAX)

Hier lässt sich gut beobachten, wie sich die Wahl des Konfidenzniveaus auf die VaR-Schätzung auswirkt. Denn wie auch bereits in Kapitel 2.1 erwähnt, wird der VaR beim Konfidenzniveau von 99% tendenziell höher geschätzt als bei dem 95%igen Konfidenzniveau. Dies wird insbesondere bei den Extrema der VaR-Schätzung deutlich.

Bei Anwendung der Varianz-Kovarianz-Methode ohne GARCH wurde für die t-Verteilung zur Schätzung der Freiheitsgrade der Befehl *fitdistr* aus *MASS* verwendet, welcher nach der Maximum-Likelihood-Schätzung vorgeht. Da es bei einigen Datensätzen Fehler mit der Optimierung gab, wurde die Funktion zur Schätzung des VaR um den Befehl *tryCatch* ergänzt. Mithilfe des *na.locf*-Befehls aus *zoo* wurde für die Werte, bei denen es zu einem Optimierungsfehler kam, die jeweils letzte Schätzung fortgeschrieben.

Beispielhaft sind in den Abbildungen 3 und 4 für ein Konfidenzniveau von 95% die Ergebnisse der Schätzung nach der Normalverteilung und t-Verteilung für den BCOM-Index zu sehen. Für das 99%ige Konfidenzniveau sind die VaR-Schätzungen des GSCI in den Abbildungen A.1 und A.2 im Anhang abgebildet.

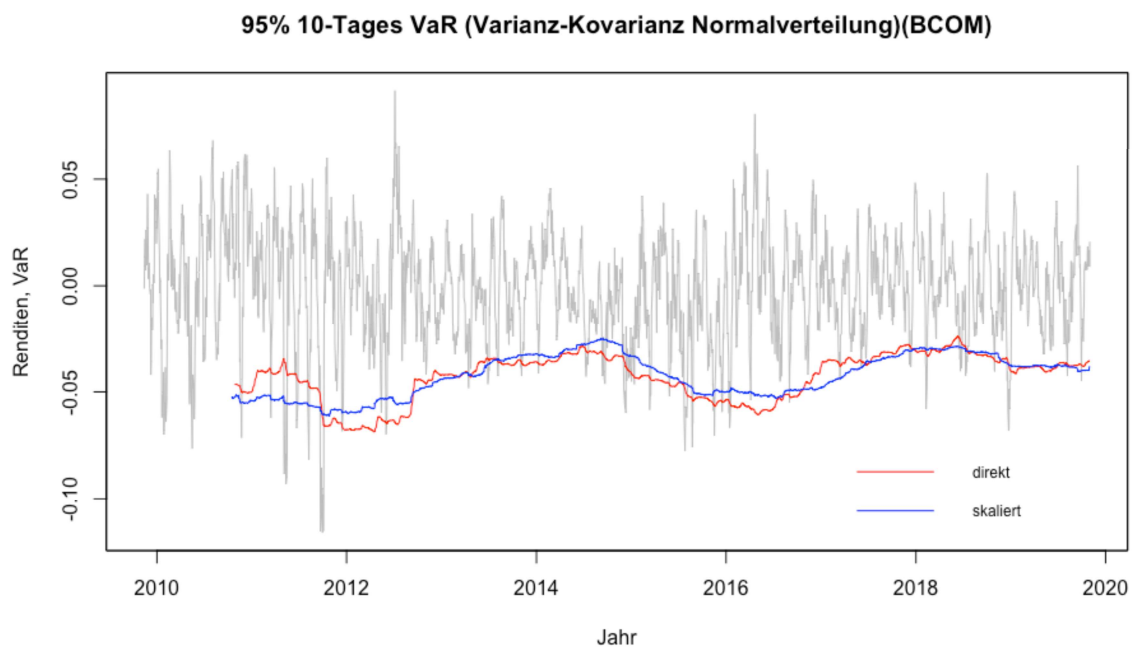

Abbildung 3: 95% 10-Tages VaR (Varianz-Kovarianz mit Normalverteilung, BCOM)

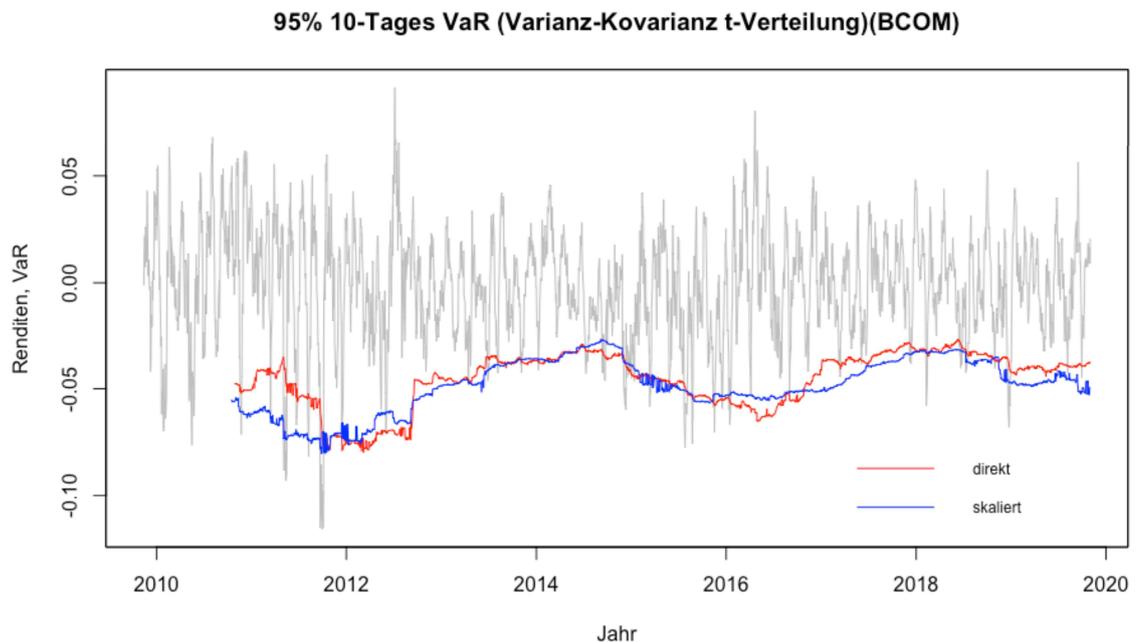

Abbildung 4: 95% 10-Tages VaR (Varianz-Kovarianz mit t-Verteilung, BCOM)

Bei Erweiterung der Varianz-Kovarianz-Methode mit dem GARCH-Modell wird in *R* das Paket *rugarch* und die Ordnung (1,1), welche generell am häufigsten Anwendung findet, verwendet (vgl. McNeil, Frey und Embrechts 2015, S. 118; Meyer 1999, S. 114). Ebenso wie die bedingte Standardabweichung werden auch die Freiheitsgrade durch das Modell geschätzt. Zusätzlich wird das GARCH-Modell um die Schätzung des bedingten Erwartungswertes mit einem ARMA-Modell ergänzt. Die Ordnung des ARMA-Modells wird dabei mit dem Befehl *auto.arima* aus *forecast* für jeden Datensatz einzeln bestimmt, indem jeweils das beste Modell gewählt wird.

Da das GARCH-Modell bei der t-Verteilung für einige Daten nicht konvergiert ist, wurde hier wieder mit dem *tryCatch*-Befehl gearbeitet und analog bei Auftreten eines Konvergenzfehlers der aktuellste Schätzwert des VaR fortgeschrieben.

Aufgrund der Rechenzeit, die bei Anwendung der GARCH-Modelle deutlich länger ist, werden als Zeitraum für die rollierende Schätzung nur die Daten der letzten 5 Jahre (31.10.2014 – 01.11.2019) verwendet.

Die Abbildungen 5 und 6 veranschaulichen die Schätzung des 95%-VaR mit dieser Methode nach den beiden Verteilungen anhand des JPYEUR-Kurses. Im Anhang sind in den Abbildungen A.3 und A.4 entsprechend die 99%-VaR-Schätzungen für den EURUSD-Kurs zu finden. Bei der t-Verteilung ist für beide Konfidenzniveaus bei der Skalierung auffällig, dass die Extrema der VaR-Schätzung deutlich ausgeprägter sind. Dies lässt sich vermutlich dadurch erklären, dass durch die t-Verteilung die Extrema der Renditeverteilung bzw. die breiten Enden auch mehr berücksichtigt werden als bei der Normalverteilung. Für die direkte VaR-Schätzung sind anhand der Abbildungen keine derart ausgeprägten Unterschiede zwischen den Verteilungen auszumachen.

Grundsätzlich ist für alle Methoden zu beobachten, dass die direkte und skalierte Schätzung voneinander abweichen. Jedoch ist hier nicht eindeutig zu erkennen, ob eine der beiden Schätzvarianten bessere bzw. genauere Schätzungen des VaR ermöglicht. Für die genauere Auswertung der Ergebnisse wird daher im Folgenden das Backtesting herangezogen.

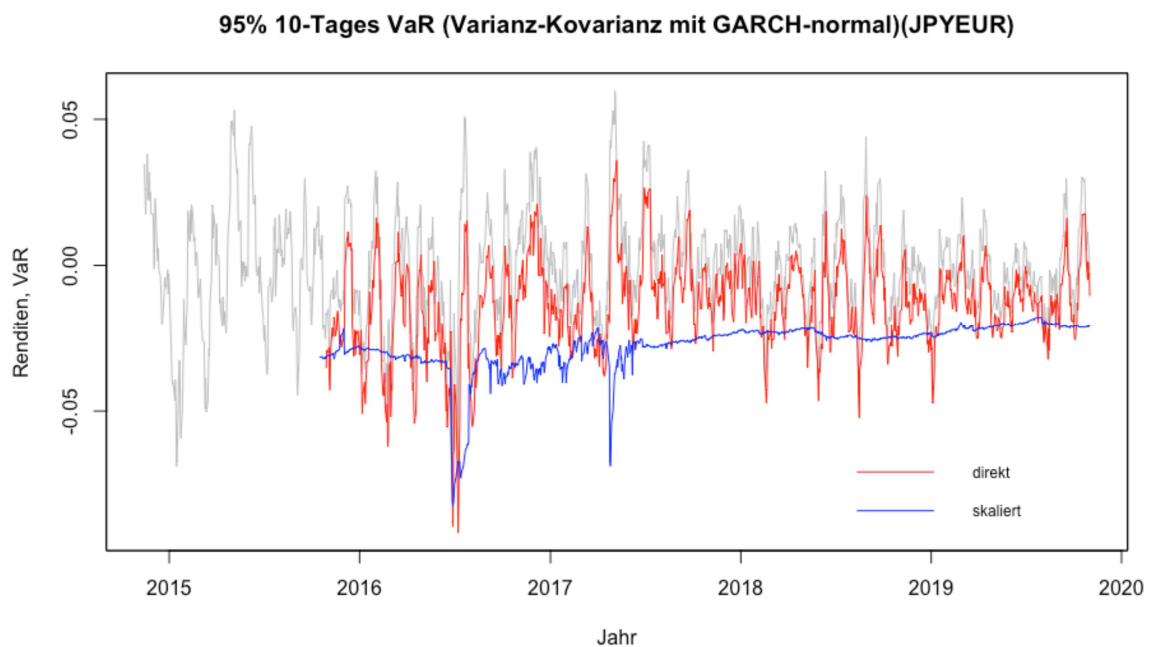

Abbildung 5: 95% 10-Tages VaR (Varianz-Kovarianz mit GARCH und Normalverteilung, JPYEUR)

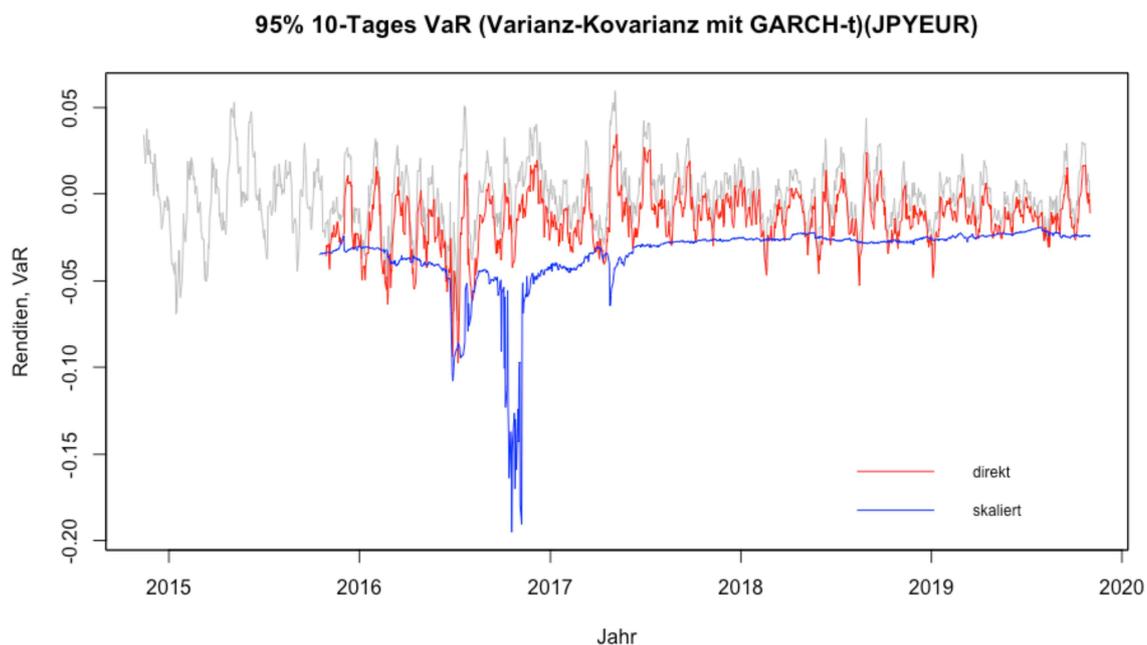

Abbildung 6: 95% 10-Tages VaR (Varianz-Kovarianz mit GARCH und t-Verteilung, JPYEUR)

### 4.3. Backtest-Ergebnisse

Für das Backtesting werden die tatsächlich beobachteten zehntägigen Renditen verwendet. Zur Anwendung des unconditional und conditional coverage Backtests wird in *R* der Befehl *VaRTest* aus dem Paket *rugarch* verwendet. Da der Backtest-Zeitraum und die VaR-Schätzungen einen identischen Zeitraum umfassen müssen, wurden diese vorher entsprechend aneinander angepasst.

Bei der Auswertung der Ergebnisse wurden die Werte der LR-Teststatistik der beiden Backtests sowie die tatsächliche Anzahl der Überschreitungen des VaR miteinbezogen. Im Folgenden wird die Teststatistik des unconditional bzw. conditional coverage Tests mit $LR_{uc}$ bzw. $LR_{cc}$ bezeichnet.

Für ein Konfidenzniveau von 95% wurde die Anzahl der Überschreitungen für alle Datensätze nach den verschiedenen VaR-Methoden in der Tabelle 3 zusammengefasst. Die erwartete Anzahl an Überschreitungen liegt bei 117 bzw. wegen des kürzeren Schätzzeitraums für die VaR-Schätzung mit den GARCH-Modellen bei 52. Anhand der erwarteten Überschreitungsanzahl wurde farblich hervorgehoben, ob eine Unterschätzung (graue Felder) oder eine Überschätzung des Risikos (blaue Felder) vorliegt. Wenn die tatsächlichen

Überschreitungen mehr als erwartet sind, wird das Risiko unterschätzt und somit der VaR tendenziell zu niedrig geschätzt. Umgekehrt gilt bei einer Überschätzung des Risikos, wo der VaR zu hoch geschätzt wird, dass weniger Überschreitungen zu beobachten sind, als erwartet werden (vgl. Peitz 2016, S. 102).

Tabelle 3: Anzahl der Überschreitungen (95% 10-Tages VaR)

| Signifikanz | 95% Erwartete Überschreitungen: 117 | | | | | | 95% Erwartete Überschreitungen: 52 | | | |
|---|---|---|---|---|---|---|---|---|---|---|
| VaR Methode | Historische Simulation | | Varianz-Kovarianz-Methode | | | | Varianz-Kovarianz mit GARCH | | | |
| | | | Normalverteilung | | t-Verteilung | | Normalverteilung | | t-Verteilung | |
| | direkt | skaliert | direkt | skaliert | direkt | skaliert | direkt | skaliert | direkt | skaliert |
| DAX | 139 | 108 | 151 | 121 | 125 | 86 | 51 | 43 | 42 | 17 |
| S&P | 123 | 87 | 143 | 91 | 94 | 53 | 43 | 19 | 26 | 7 |
| GSCI | 133 | 172 | 162 | 183 | 145 | 97 | 49 | 69 | 39 | 31 |
| BCOM | 133 | 167 | 147 | 168 | 133 | 113 | 46 | 58 | 40 | 42 |
| EURUSD | 107 | 61 | 89 | 56 | 71 | 33 | 37 | 27 | 28 | 13 |
| JPYEUR | 106 | 103 | 88 | 106 | 64 | 65 | 45 | 45 | 38 | 31 |

Überschätzung des Risikos    Unterschätzung des Risikos

Tabelle 4: Teststatistik nach unconditional coverage Backtest (95% 10-Tages VaR)

| Signifikanz | 95% Kritischer Wert: 3,841 | | | | | | | | | |
|---|---|---|---|---|---|---|---|---|---|---|
| VaR Methode | Historische Simulation | | Varianz-Kovarianz-Methode | | | | Varianz-Kovarianz mit GARCH | | | |
| | | | Normalverteilung | | t-Verteilung | | Normalverteilung | | t-Verteilung | |
| | direkt | skaliert | direkt | skaliert | direkt | skaliert | direkt | skaliert | direkt | skaliert |
| DAX | 3,883 | 0,847 | 9,199 | 0,103 | 0,481 | 9,819 | 0,037 | 1,867 | 2,304 | 33,701 |
| S&P | 0,257 | 9,177 | 5,419 | 6,845 | 5,337 | 46,568 | 1,867 | 29,292 | 16,999 | 64,568 |
| GSCI | 2,040 | 23,326 | 15,868 | 32,977 | 6,276 | 4,029 | 0,230 | 5,090 | 3,916 | 10,669 |
| BCOM | 2,040 | 19,436 | 7,193 | 20,188 | 2,040 | 0,192 | 0,844 | 0,621 | 3,327 | 2,304 |
| EURUSD | 1,036 | 34,539 | 7,964 | 41,786 | 22,511 | 88,492 | 5,255 | 15,586 | 14,249 | 44,019 |
| JPYEUR | 1,244 | 1,988 | 8,559 | 1,244 | 30,600 | 29,351 | 1,138 | 1,138 | 4,558 | 10,669 |

Nulhypothese abgelehnt

In der Tabelle 4 sind die Ergebnisse der $LR_{uc}$-Teststatistik aufgelistet. Die roten Felder kennzeichnen die Modelle, bei denen die Nullhypothese abgelehnt wird, da die Teststatistik den kritischen Wert von 3,841 überschreitet. Das bedeutet, dass die Anzahl der Überschreitungen dort zu sehr von der erwarteten Anzahl abweicht. In Kombination mit den Ergebnissen aus Tabelle 3 kann geschlussfolgert werden, ob das Risiko in diesen Fällen über- oder unterschätzt wird.

Für die historische Simulation fällt bei der Betrachtung der Tabellen 3 und 4 auf, dass zur Schätzung des VaR die Skalierung für die Rohstoffindizes ungeeignet ist, da das Risiko unterschätzt wird. Auch für die Währungskurse kann mit der direkten Ermittlung der VaR in der Hinsicht genauer geschätzt werden, dass die Anzahl der Überschreitungen näher an der erwarteten Anzahl liegt. Dies lässt sich ebenfalls anhand des kleineren $LR_{uc}$-Wertes bei der direkten Ermittlung feststellen. Durch die Skalierung wird allerdings für die Währungskurse und auch für den Datensatz S&P das Risiko überschätzt. Nur für den DAX würde die Auswahl auf das skalierte Modell fallen.

Bei der Varianz-Kovarianz Methode mit der Normalverteilung ist die Skalierung für die Rohstoffindizes wieder ungeeignet. Wie bei der historischen Simulation wird das Risiko unterschätzt. Jedoch sollte hier darauf hingewiesen werden, dass für alle Datensätze bei der direkten Schätzung des VaR die Anzahl der Überschreitungen signifikant von der erwarteten Anzahl abweicht. Für den DAX und JPYEUR-Kurs sind stattdessen die Ergebnisse bei Anwendung der Skalierung deutlich besser.

Für die t-Verteilung wird die Skalierung bei den Aktienindizes und Währungskursen abgelehnt. Aber auch bei dieser Verteilung wird bei der direkten Schätzung des VaR mehrheitlich die Nullhypothese des unconditional coverage Backtests abgelehnt. Nichtsdestotrotz lässt sich insbesondere für die Aktien- und Rohstoffindizes durch die Verwendung der t-Verteilung eine Besserung der $LR_{uc}$-Werte erkennen. Für den BCOM lässt sich anhand des kleineren $LR_{uc}$-Wertes erkennen, dass die Skalierung bessere Schätzergebnisse als die direkte Ermittlung liefert.

Durch die Erweiterung der Varianz-Kovarianz-Methode mit dem GARCH(1,1)-Modell zeigt sich bei der Normalverteilung für die direkte Ermittlung verglichen zur Schätzung ohne GARCH eine deutliche Verbesserung der $LR_{uc}$-Werte. Das heißt, für die meisten Datensätze entsprechen nun die tatsächlichen Überschreitungen der Erwartung. Für die Mehrheit der Daten liefert die direkte Ermittlung bessere Schätzergebnisse als die Skalierung.

Bei der t-Verteilung wird nach der Skalierung bei fast allen Datensätzen der VaR signifikant überschätzt. Nur bei dem BCOM-Index zeigt die Skalierung hinsichtlich der Überschreitungsanzahl knapp bessere Ergebnisse als die direkte Ermittlung.

Hinsichtlich der Ergebnisse des conditional coverage Backtests in Tabelle 5 ist zu sehen, dass bei der historischen Simulation und der Varianz-Kovarianz-Methode ohne GARCH die Nullhypothese für alle Datensätze abgelehnt wird. Dies gilt sowohl für die skalierte als auch für die direkte VaR-Schätzung. Somit sind bei den nach dem unconditional coverage Test akzeptierten Modellen die Überschreitungen nicht unabhängig und damit insgesamt nur unzureichend zur Schätzung des VaR geeignet.

Dahingegen kann für die Varianz-Kovarianz-Methode mit GARCH bei der Normalverteilung festgestellt werden, dass aufgrund der Unabhängigkeit der Überschreitungen die direkte Ermittlung zu bevorzugen ist. Für die Modelle, bei denen mit der Skalierung die Überschreitungsanzahl nahe der erwarteten lag, sind die Überschreitungen abhängig.

Für die t-Verteilung ist das Ergebnis nicht ganz so eindeutig wie bei der Normalverteilung. Grundsätzlich liegen aber bei der direkten Ermittlung deutlich kleinere $LR_{cc}$-Werte vor als bei der Skalierung, wo für alle Datensätze die Nullhypothese abgelehnt wird.

Tabelle 5: Teststatistik nach conditional coverage Backtest (95% 10-Tages VaR)

| Signifikanz | 95% Kritischer Wert: 5,991 | | | | | | | | | |
|---|---|---|---|---|---|---|---|---|---|---|
| VaR Methode | Historische Simulation | | Varianz-Kovarianz-Methode | | | | Varianz-Kovarianz mit GARCH | | | |
| | | | Normalverteilung | | t-Verteilung | | Normalverteilung | | t-Verteilung | |
| | direkt | skaliert | direkt | skaliert | direkt | skaliert | direkt | skaliert | direkt | skaliert |
| DAX | 358,120 | 324,380 | 434,997 | 371,519 | 338,980 | 307,582 | 0,148 | 104,946 | 4,862 | 63,684 |
| S&P | 478,922 | 363,447 | 564,552 | 406,184 | 400,080 | 297,590 | 4,187 | 63,605 | 18,960 | 69,059 |
| GSCI | 414,292 | 551,980 | 584,523 | 591,314 | 504,568 | 373,673 | 0,274 | 158,584 | 4,115 | 66,897 |
| BCOM | 370,187 | 553,442 | 477,105 | 549,410 | 423,445 | 340,341 | 5,077 | 136,349 | 6,509 | 92,454 |
| EURUSD | 293,227 | 214,013 | 284,762 | 218,317 | 234,001 | 179,363 | 5,340 | 72,926 | 14,330 | 82,121 |
| JPYEUR | 279,916 | 345,348 | 298,433 | 314,463 | 247,155 | 252,034 | 1,140 | 114,375 | 4,824 | 99,472 |

Nulhypothese abgelehnt

Für den 99%-VaR sind die Ergebnisse in den Tabellen 6 bis 8 zusammengefasst. Jetzt liegt im Gegensatz zum Konfidenzniveau von 95% bei der historischen Simulation nach der Skalierung deutlich öfter eine korrekte Anzahl an Überschreitungen vor. Verglichen mit der direkten Schätzung sind die Ergebnisse mehrheitlich für die Skalierung besser. Insbesondere bei den Rohstoffindizes liefert die Skalierung deutlich bessere Ergebnisse bei der Überschreitungsanzahl als die direkte Ermittlung, was bei 95%iger Signifikanz genau umgekehrt war. Grundsätzlich lässt sich hier für alle Datensätze beobachten, dass bei der Skalierung im Vergleich zur direkten Ermittlung weniger Überschreitungen auftreten, also der VaR tendenziell höher geschätzt wird.

Bei der Varianz-Kovarianz Methode ohne GARCH-Modell ist für die Währungskurse bei der Normalverteilung die direkte Schätzung des VaR eindeutig besser. Jedoch kann anhand der beiden Datensätze keine eindeutige Tendenz zur Über- oder Unterschätzung des Risikos durch die Skalierung ausgemacht werden. Auch bei den Rohstoffindizes ist die direkte Ermittlung besser geeignet, in diesem Fall aber für die t-Verteilung. Außerdem lässt sich beobachten, dass, ähnlich wie bei der historischen Simulation, bei der t-Verteilung das Risiko im Vergleich zur direkten Ermittlung überschätzt wird.

Tabelle 6: Anzahl der Überschreitungen (99% 10-Tages VaR)

| Signifikanz | 99% Erwartete Überschreitungen: 23 | | | | | | 99% Erwartete Überschreitungen: 10 | | | |
|---|---|---|---|---|---|---|---|---|---|---|
| VaR Methode | Historische Simulation | | Varianz-Kovarianz-Methode | | | | Varianz-Kovarianz mit GARCH | | | |
| | | | Normalverteilung | | t-Verteilung | | Normalverteilung | | t-Verteilung | |
| | direkt | skaliert | direkt | skaliert | direkt | skaliert | direkt | skaliert | direkt | skaliert |
| DAX | 41 | 30 | 54 | 46 | 35 | 16 | 11 | 9 | 7 | 1 |
| S&P | 28 | 16 | 61 | 35 | 21 | 6 | 17 | 1 | 6 | 0 |
| GSCI | 36 | 24 | 49 | 62 | 28 | 11 | 14 | 9 | 7 | 1 |
| BCOM | 40 | 22 | 39 | 42 | 22 | 7 | 11 | 5 | 8 | 2 |
| EURUSD | 31 | 10 | 19 | 13 | 4 | 3 | 8 | 3 | 6 | 3 |
| JPYEUR | 29 | 23 | 28 | 32 | 6 | 6 | 11 | 8 | 7 | 2 |

Überschätzung des Risikos   Unterschätzung des Risikos

Tabelle 7: Teststatistik nach unconditional coverage Backtest (99% 10-Tages VaR)

| Signifikanz | 99% Kritischer Wert: 6,635 | | | | | | | | | |
|---|---|---|---|---|---|---|---|---|---|---|
| VaR Methode | Historische Simulation | | Varianz-Kovarianz-Methode | | | | Varianz-Kovarianz mit GARCH | | | |
| | | | Normalverteilung | | t-Verteilung | | Normalverteilung | | t-Verteilung | |
| | direkt | skaliert | direkt | skaliert | direkt | skaliert | direkt | skaliert | direkt | skaliert |
| DAX | 10,741 | 1,659 | 29,203 | 16,970 | 4,921 | 2,736 | 0,027 | 0,219 | 1,315 | 14,329 |
| S&P | 0,812 | 2,736 | 41,915 | 4,921 | 0,283 | 18,778 | 3,461 | 14,329 | 2,278 | 21,045 |
| GSCI | 5,755 | 0,010 | 21,249 | 43,871 | 0,812 | 8,388 | 1,087 | 0,219 | 1,315 | 14,329 |
| BCOM | 9,639 | 0,101 | 8,589 | 11,892 | 0,101 | 16,190 | 0,027 | 3,578 | 0,641 | 10,388 |
| EURUSD | 2,184 | 10,013 | 0,939 | 5,672 | 25,031 | 28,865 | 0,641 | 7,494 | 2,278 | 7,494 |
| JPYEUR | 1,201 | 0,012 | 0,812 | 2,776 | 18,778 | 18,778 | 0,027 | 0,641 | 1,315 | 10,388 |

Nulhypothese abgelehnt

Tabelle 8: Teststatistik nach conditional coverage Backtest (99% 10-Tages VaR)

| Signifikanz | 99% Kritischer Wert: 9,210 | | | | | | | | | |
|---|---|---|---|---|---|---|---|---|---|---|
| VaR Methode | Historische Simulation | | Varianz-Kovarianz-Methode | | | | Varianz-Kovarianz mit GARCH | | | |
| | | | Normalverteilung | | t-Verteilung | | Normalverteilung | | t-Verteilung | |
| | direkt | skaliert | direkt | skaliert | direkt | skaliert | direkt | skaliert | direkt | skaliert |
| DAX | 167,529 | 155,867 | 253,407 | 230,956 | 140,013 | 89,584 | 2,720 | 51,219 | 1,409 | 14,331 |
| S&P | 116,022 | 89,584 | 312,222 | 150,840 | 105,584 | 48,131 | 8,303 | 14,331 | 7,417 | 21,045 |
| GSCI | 137,558 | 94,972 | 191,549 | 288,210 | 94,724 | 61,728 | 1,467 | 38,928 | 1,409 | 14,331 |
| BCOM | 102,476 | 90,158 | 161,827 | 197,485 | 101,704 | 31,736 | 0,260 | 9,505 | 0,764 | 20,749 |
| EURUSD | 88,440 | 31,600 | 89,238 | 53,459 | 33,566 | 28,873 | 0,764 | 15,768 | 2,347 | 15,768 |
| JPYEUR | 73,381 | 98,180 | 116,022 | 160,484 | 48,131 | 48,131 | 0,260 | 20,635 | 1,409 | 10,395 |

Nulhypothese abgelehnt

Mit den GARCH-Modellen ist für beide Verteilungen zu erkennen, dass die direkte Ermittlung eine korrektere Schätzung des VaR hinsichtlich beider Backtest-Methoden ermöglicht. Durch die Skalierung wird insbesondere für die t-Verteilung der VaR signifikant überschätzt. Diese Tendenz zur Überschätzung ist auch bei der Normalverteilung zu beobachten, jedoch ist die Überschreitungsanzahl häufig noch angemessen. Allerdings sind die Überschreitungen nicht unabhängig, wie es dem conditional coverage Backtest zu entnehmen ist.

# 5. Diskussion

Für die Mehrheit der Methoden kann den Ergebnissen aus Kapitel 4.3 entnommen werden, dass die Skalierung mit der Quadratwurzel-T-Regel meist den VaR überschätzt. Besonders bei einem Konfidenzniveau von 99% ist dies im Vergleich mit der direkten Ermittlung zu sehen.

Dass die Skalierung den VaR hingegen unterschätzt (vgl. Daníelsson und Zigrand 2003, S. 1; Wang, Yeh und Cheng 2011, S. 1158), lässt sich nur eindeutig für die Rohstoffindizes bei Anwendung der historischen Simulation und Varianz-Kovarianz-Methode mit Normalverteilung bei einem Konfidenzniveau von 95% beobachten.

Bis auf vereinzelte Ausnahmen sind für beide Signifikanzniveaus und die meisten Methoden bei der direkten Ermittlung bessere Schätzergebnisse zu sehen. Somit stimmen diese Ergebnisse mit den diversen Untersuchungen in der Hinsicht überein, dass die Skalierung keine guten Schätzungen des VaR ermöglicht (vgl. BCBS 2011, S. 1). Allerdings zeigt die Skalierung bei der historischen Simulation für ein Konfidenzintervall von 99% größtenteils bessere Ergebnisse.

Wie auch Meyer (1999, S. 363 f.) anmerkte, lässt sich ebenso in den Ergebnissen dieser Arbeit nicht immer eine allgemeine oder eindeutige Aussage zur Über- oder Unterschätzung des Risikos durch die Skalierung treffen. Vor allem für die historische Simulation und die Varianz-Kovarianz-Methode ohne GARCH-Modell ist dies zu beobachten. Oftmals sind je nach Datenkategorie und auch teilweise innerhalb der Datenkategorie unterschiedliche Ergebnisse zu erkennen.

Für das GARCH(1,1)-Modell kann unter der Annahme der Normalverteilung grundsätzlich den Ergebnissen von Diebold et al. (1998, S. 105) zugestimmt werden. Bei der Mehrheit der Daten zeigt die Skalierung angemessene Ergebnisse zur Schätzung des Risikos. Dies gilt jedoch nicht für die t-Verteilung, denn dort wird insbesondere bei einem höheren Konfidenzniveau der VaR durch die Skalierung überschätzt. Tendenziell ist auch für die Normalverteilung im Vergleich zur direkten Ermittlung zu beobachten, dass die Skalierung den VaR höher schätzt. In dieser Hinsicht muss den Ergebnissen von Saadi und Rahman (2008, S. 272) und Cameron et al. (2016, S. 78) widersprochen werden, welche eine Unterschätzung der bedingten Volatilität durch die Skalierung ergaben.

Bei der Anwendung des GARCH-Modells lässt sich insgesamt sagen, dass die direkte Ermittlung bessere Ergebnisse als die skalierte Schätzung ermöglicht. Für ein höheres Konfidenzniveau wird dies umso deutlicher.

Beim Vergleich aller untersuchten VaR-Methoden miteinander fällt hinsichtlich der Überschreitungsanzahl und Unabhängigkeit der Überschreitungen die Auswahl auf die direkte Schätzung mit der Varianz-Kovarianz-Methode mit GARCH. Da durch die GARCH-Modelle in der VaR-Schätzung die Zeitabhängigkeit der Volatilität mit einbezogen wird, sind die Überschreitungen auch in den meisten Fällen unabhängig (vgl. Jorion 2011, S. 362). Bei der Skalierung ist das aber nicht der Fall. Ein Grund dafür könnte sein, dass die eintägige Volatilität sich in ihrem Verlauf und auch den Parametern zur Schätzung von der zehntägigen unterscheidet.

Als Verteilungsannahme, die bei dieser VaR-Schätzmethode die besseren Ergebnisse ermöglicht, sind je nach Konfidenzniveau unterschiedliche Ergebnisse festzustellen. Für ein Signifikanzniveau von 95% sind für alle Datenkategorien bessere Ergebnisse bei der Normalverteilung zu beobachten. Bei dem Konfidenzniveau von 99% ist für die Währungskurse immer noch die Normalverteilung besser, hingegen für die Aktienindizes nun die t-Verteilung. Denn wie es der deskriptiven Statistik in Kapitel 4.1 zu entnehmen ist, weisen insbesondere die 10-Tages Renditen der Aktienkurse deutlich höhere Kurtosis-Werte auf als für die Normalverteilung üblich. Im Gegensatz dazu ähneln bei den Währungskursen die Schiefe und Kurtosis der 10-Tages Renditen fast der Normalverteilung.

Weiterführend gibt es noch die Möglichkeit, die Quadratwurzel-T-Regel für andere bzw. längere Haltedauern als 10 Tage zu testen. Für eine Haltedauer von 30 Tagen kann hierfür auf die Arbeit von Wang et al. (2011) verwiesen werden.

Zusätzlich kann bei der rollierenden VaR-Schätzung die Fenstergröße auf z.B. 500 oder 1000 Beobachtungen variiert werden, wie es in den Untersuchungen von Kuester et al. (2006) zu finden ist.

Außerdem könnte die Skalierung auch mit anderen Backtest-Methoden, die nicht auf der Überschreitungsanzahl basieren, getestet werden.

# 6. Fazit

Zusammenfassend lässt sich sagen, dass die direkte Schätzung des VaR verglichen mit der Skalierung über die Quadratwurzel-T-Regel größtenteils bessere Schätzergebnisse ermöglicht. Gleichzeitig weist aber für einige Datensätze und Methoden die Skalierung auch angemessene oder teilweise bessere Ergebnisse als die direkte VaR-Schätzung auf. Grundsätzlich ist die Eignung der Skalierung also von der verwendeten VaR-Schätzmethode und dem Datensatz abhängig.

Ein sehr eindeutiges Ergebnis ist jedoch für die Varianz-Kovarianz-Methode mit dem GARCH-Modell zu erkennen. Durch die Skalierung wird der VaR überschätzt, während die direkte Ermittlung bessere Schätzergebnisse ermöglicht. Vor allem sind die Überschreitungen bei der direkten Schätzung unabhängig, da die Veränderung der Volatilität in der Zeit der Haltedauer entsprechend richtig berücksichtigt wird.

Insgesamt sollte die Erlaubnis zur Verwendung der Quadratwurzel-T-Regel in den Regulierungsvorschriften überdacht werden. Zwar ist anhand der Ergebnisse dieser Arbeit zu erkennen, dass die Skalierung auch gute und korrekte VaR-Schätzungen ermöglichen kann. Deutlich wichtiger ist jedoch wohl in Betracht zu ziehen, dass, je nachdem, um welche Methoden oder welchen Datensatz es sich handelt, es auch zu erheblichen Konsequenzen für die Risikovorsorge kommen kann. Besonders, da bei der Skalierung nicht immer vermieden wird, dass in einem Zeitabschnitt gehäuft Verluste auftreten, die die VaR-Schätzung überschreiten.

# Anhang

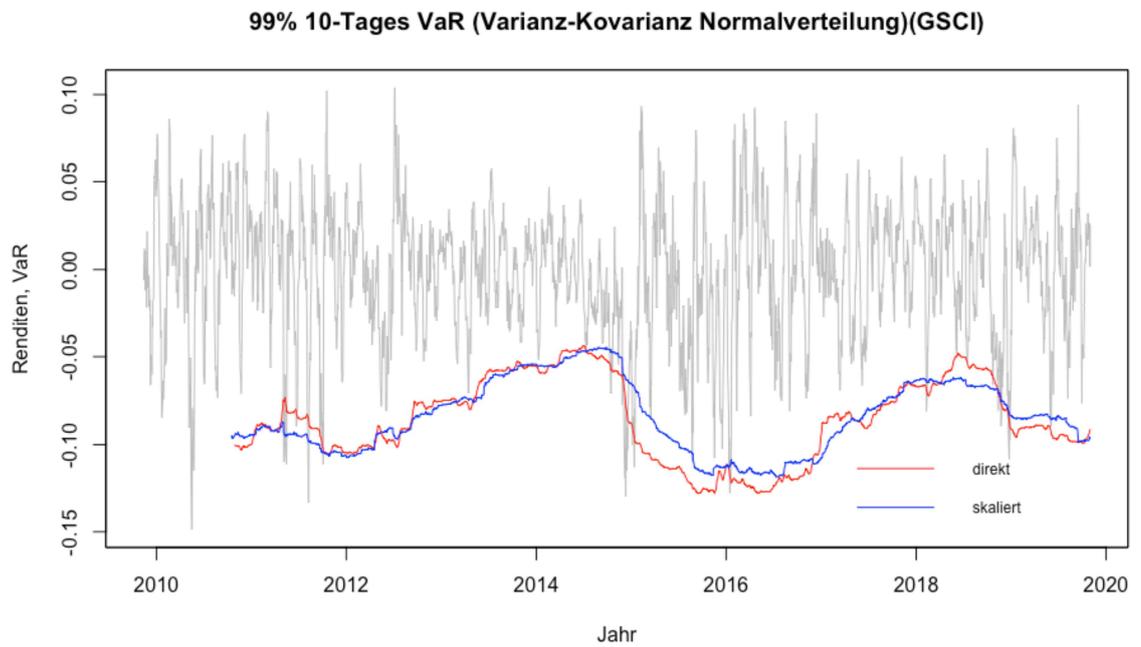

Abbildung A.1: 99% 10-Tages VaR (Varianz-Kovarianz mit Normalverteilung, GSCI)

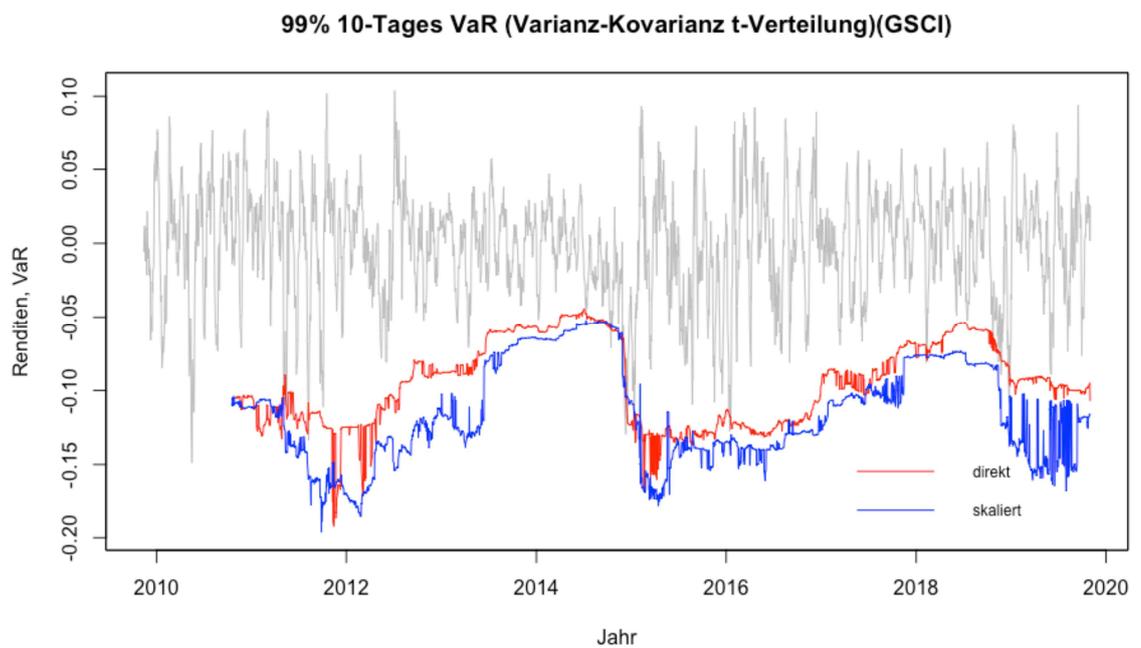

Abbildung A.2: 99% 10-Tages VaR (Varianz-Kovarianz mit t-Verteilung, GSCI)

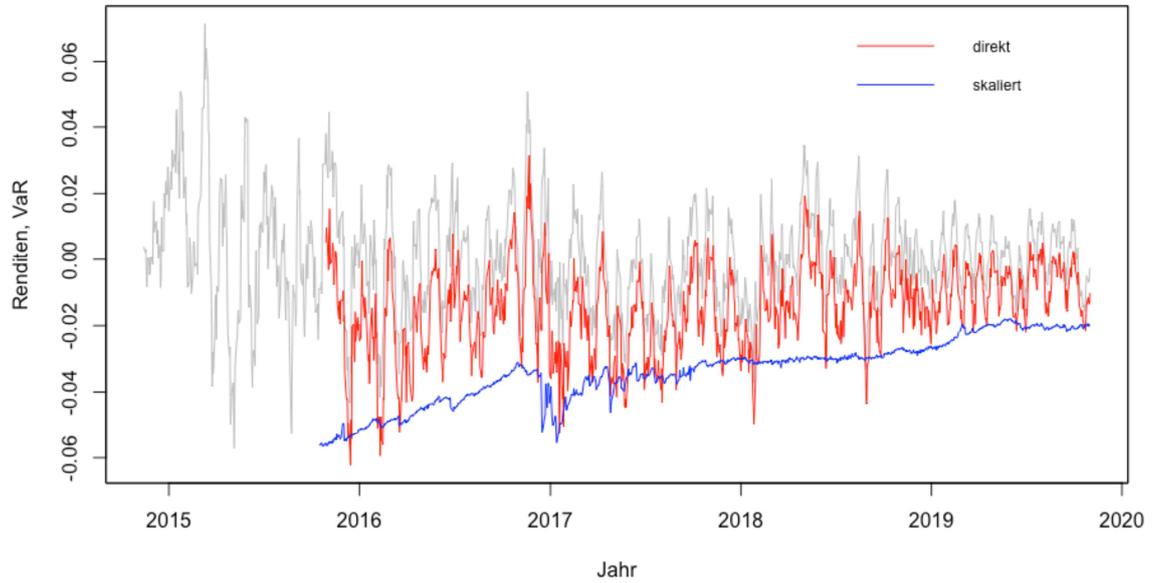

Abbildung A.3: 99% 10-Tages VaR (Varianz-Kovarianz mit GARCH und Normalverteilung, EURUSD)

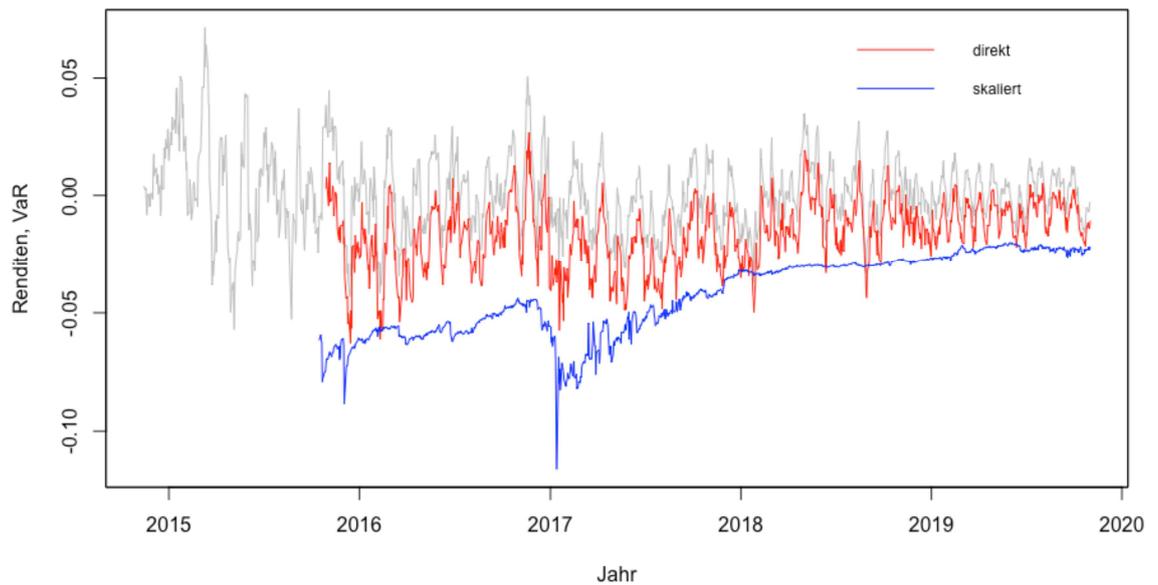

Abbildung A.4: 99% 10-Tages VaR (Varianz-Kovarianz mit GARCH und t-Verteilung, EURUSD)

# Literaturverzeichnis

# Ehrenwörtliche Erklärung

Ich versichere durch eigenhändige Unterschrift, dass ich die vorliegende Arbeit selbständig und ohne unerlaubte Hilfe Dritter angefertigt habe. Alle Stellen, die inhaltlich oder wörtlich aus Veröffentlichungen stammen, sind kenntlich gemacht. Diese Arbeit lag nach meinem Informationsstand in gleicher oder ähnlicher Weise noch keiner Prüfungsbehörde vor und wurde bisher noch nicht veröffentlicht. Ich bin mir darüber bewusst, dass bei Abgabe einer falschen Erklärung die Prüfung als nicht bestanden gilt. Im dringenden Verdachtsfall kann meine Arbeit unter Zuhilfenahme des Dienstes „Turnitin" geprüft werden. Dabei erlaube ich [ ] die Ablage meiner Arbeit im institutsinternen Speicher / [ ] keine Ablage meiner Arbeit. Unabhängig vom Ergebnis der Prüfung durch „Turnitin" wird immer eine individuelle Prüfung und Bewertung der Arbeit vorgenommen. Darüber hinaus wird der Inhalt der Arbeit Dritten nicht ohne meine ausdrückliche Genehmigung zugänglich gemacht.

Paderborn, den 21.01.2020 \_\_\_\_\_\_\_\_\_\_\_\_\_\_\_\_\_\_\_\_\_\_\_\_\_\_\_\_\_\_\_\_\_\_\_\_\_

Unterschrift